\documentclass[12pt]{article}
\usepackage{amsfonts, amssymb, amscd}
\usepackage{cite}

\begin{document}

\title{G\"{o}del-Type  Metrics in Einstein-Aether Theory II: Nonflat Background in Arbitrary Dimensions}
\author{ Metin G\"{u}rses$^{(a),(b)}$\footnote{gurses@fen.bilkent.edu.tr} and \c{C}etin \c{S}ent\"{u}rk$^{(b)}$\footnote{cetin.senturk@bilkent.edu.tr} \\
{\small (a) Department of Mathematics, Faculty of Sciences}\\
{\small Bilkent University, 06800 Ankara, Turkey}\\
 {\small (b) Department of Physics, Faculty of Sciences}\\
{\small Bilkent University, 06800 Ankara, Turkey}
}

\begin{titlepage}
\maketitle
\thispagestyle{empty}

\begin{abstract}
It was previously proved that the G\"{o}del-type metrics with flat three-dimensional background metric solve exactly the field equations of the Einstein-Aether theory in four dimensions. We generalize this result by showing that the stationary G\"{o}del-type metrics with nonflat background in $D$ dimensions solve exactly the field equations of the Einstein-Aether theory. The reduced field equations are the $(D-1)$-dimensional Euclidean Ricci-flat and the $(D-1)$-dimensional source-free Maxwell equations, and the parameters of the theory are left free except $c_{1}-c_{3}=1$. We give a method to produce exact solutions of the Einstein-Aether theory from the G\"{o}del-type metrics in $D$ dimensions. By using this method, we present explicit exact solutions to the theory by considering the particular cases: ($D-1$)-dimensional Euclidean flat, conformally flat, and Tangherlini backgrounds.

\end{abstract}
\end{titlepage}


\section{Introduction}

G\"{o}del-type metrics were introduced in \cite{gur1,gur2} as a generalization of the usual G\"{o}del solution \cite{godel} in general relativity and used to obtain new solutions to various (super)gravity theories in diverse dimensions. Just like in the original G\"{o}del solution, the spacetimes described by such metrics also contain closed timelike and closed null curves when the characteristic vector field that defines a G\"{o}del-type metric is a Killing vector \cite{gur3}. In three dimensions, G\"{o}del type of metrics naturally provide charged perfect fluid solutions and also give some special solutions to Topologically Massive Gravity \cite{gur4}.

In a recent publication \cite{gur}, it was shown that the G{\" o}del type of metrics with flat backgrounds form an exact solution to the field equations of the Einstein-Aether theory \cite{jm,jac} in four dimensions. The parameters of the theory corresponding to this solution obey the constraint $c_{1}-c_{3}=1$ which decouples the twist degree of freedom from the Lagrangian. Here we generalize this work by extending the dimension to $D(\ge 4$) and relaxing the assumption that $h_{\mu \nu}$ be a flat metric of a ($D-1$)-dimensional (Euclidean) space. We shall use the metric signature $(-,+,+,+,\ldots)$ throughout the paper.

Let $(M,g)$ be a $D$-dimensional spacetime geometry with the G{\"o}del-type metric tensor [1,2,4]
\begin{equation}\label{godel}
g_{\mu \nu}=h_{\mu \nu}-u_{\mu}\, u_{\nu},
\end{equation}
where $h_{\mu \nu}$ is the metric of a ($D-1$)-dimensional locally Euclidean Einstein space, which we sometimes call the \textquotedblleft background,"  and $u^{\mu}$ is a timelike unit vector field ($u^{\mu}\, u_{\mu}=-1$). Here we take $u^{\mu}=-\frac{1}{u_{0}}\, \delta_{0}^{\mu}$ without loosing any generality and assume $u^{\mu}h_{\mu \nu}=0$. We also assume that the background metric is time-independent, i.e., $\partial_{0}h_{\mu \nu}=0$. The inverse of $g_{\mu \nu}$ is given by
\begin{equation}
g^{\mu \nu}=\bar{h}^{\mu \nu}+(-1+\bar{h}^{\alpha \beta}\,u_{\alpha} u_{\beta}) u^{\mu} u^{\nu}+(\bar{h}^{\nu \alpha} u_{\alpha}) u^{\mu}+
(\bar{h}^{\mu \alpha} u_{\alpha}) u^{\nu},
\end{equation}
where $\bar{h}^{\mu \nu}$ is the $(D-1)$-dimensional inverse of $h_{\mu \nu}$; i.e.,
\[
\bar{h}^{\mu \alpha}\, h_{\alpha \nu}=\bar{\delta}^{\mu}_{\nu} \equiv \delta^{\mu}_{\nu}+u^{\mu}\, u_{\nu}.
\]
Defining the Christoffel symbols of $h_{\mu\nu}$ as
\begin{equation}
\gamma^{\mu}_{\alpha \beta}=\frac{1}{2}\, \bar{h}^{\mu \rho} \left(h_{\alpha \rho,\beta}+h_{\beta \rho,\alpha}-h_{\alpha \beta, \rho} \right),
\end{equation}
one can show that the Christoffel symbols of the full metric are
\begin{equation}
\Gamma^{\mu}_{\alpha \beta} = \gamma^{\mu}_{\alpha \beta} +
u^{\mu} \, u_{\sigma} \, \gamma^{\sigma}_{\alpha\beta} +
\frac{1}{2}\, (u_{\alpha} \, f^{\mu}\,_{\beta} + u_{\beta} \,
f^{\mu}\,_{\alpha}) - \frac{1}{2} \, u^{\mu} \, (u_{\alpha, \,
\beta} + u_{\beta, \, \alpha}) \, , \label{chris}
\end{equation}
where $f_{\alpha \beta} \equiv u_{\beta, \,
\alpha} - u_{\alpha, \, \beta}$ and a comma denotes partial differentiation. We assume that the indices of $u_{\mu}$ and $f_{\alpha\beta}$ are raised and lowered by the metric $g_{\mu\nu}$. By using a vertical stroke to denote covariant derivative with respect to the Christoffel symbols $\gamma^{\mu}_{\alpha \beta}$ of $h_{\mu\nu}$ so that
\begin{equation}
u_{\alpha \vert \beta} = u_{\alpha, \, \beta} -
\gamma^{\nu}_{\alpha\beta} \, u_{\nu} \, , \label{kare}
\end{equation}
the equation (\ref{chris}) can be simply written as
\begin{equation}
\Gamma^{\mu}_{\alpha \beta} = \gamma^{\mu}_{\alpha \beta} +
\frac{1}{2}\, (u_{\alpha} \, f^{\mu}\,_{\beta} + u_{\beta} \,
f^{\mu}\,_{\alpha}) - \frac{1}{2} \, u^{\mu} \, (u_{\alpha \vert
\beta} + u_{\beta \vert \alpha}) \, . \label{tilchr}
\end{equation}
We shall denote the covariant derivative with respect to the Christoffel
symbols $\Gamma^{\mu}_{\alpha \beta}$ of $g_{\mu \nu}$ by
$\nabla_{\mu}$ or by a semicolon. For instance,
\begin{equation}\label{covder}
\nabla_{\mu}\, u_{\nu} \equiv u_{\nu; \mu}= u_{\nu,
\mu}-\Gamma^{\gamma}_{\nu\mu}\, u_{\gamma}.
\end{equation}

In this work we have three main assumptions: (i) $u_{0}$ is a constant and taken to be unity, (ii) the acceleration of $u^{\mu}$ vanishes, i.e. $\dot{u}_{\mu} \equiv u^{\alpha}\, u_{\mu;\alpha}=0$, and (iii) $u^{\mu}h_{\mu \nu}=0$ and $\partial_{0}h_{\mu \nu}=0$. The first assumption eliminates the dilaton field $\phi\equiv\ln |u_{0}|$ in the field equations [2]. From the last two assumptions, we obtain that (a) $u_{\mu}$ is a Killing vector and (b) $\partial_{0} u_{\mu}=0$.

\vspace{0.5cm}
\noindent
{\bf  Proof of (a)}: We defined $f_{\alpha \beta} \equiv u_{\beta,\alpha}-u_{\alpha,\beta}=u_{\beta;\alpha}-u_{\alpha;\beta}$. Since $u^\mu$ is a unit vector, which gives $u^{\mu}\, u_{\mu;\alpha}=0$, by assumption (ii) we find
\[
u^{\mu}\, f_{\mu\alpha}=u^{\mu}\,u_{\alpha;\mu}-u^{\mu}\, u_{\mu;\alpha}=u^{\mu}\,u_{\alpha;\mu}=0.
\]
Hence it follows that
\[
u_{\alpha;\beta}=u_{\alpha, \beta}-\Gamma^{\rho}_{\alpha \beta}\,u_{\rho}=u_{\alpha,\beta}-\frac{1}{2}(u_{\alpha,\beta}+u_{\beta, \alpha})=\frac{1}{2}(u_{\alpha,\beta}-u_{\beta, \alpha})=\frac{1}{2}\, f_{\beta \alpha}.
\]
This last result tells us that $u^{\mu}$ is a Killing vector.

\vspace{0.5cm}
\noindent
{\bf  Proof of (b)}: The last two assumptions above imply that $u^{\mu}\,u_{\alpha;\mu}=u^{\mu}\,u_{\alpha,\mu}-\Gamma^{\rho}_{\alpha \mu}\,u_{\rho} u^{\mu}=0 \Rightarrow u^{\mu}\,u_{\alpha,\mu}=0$ with the choice $u^{\mu}=-\frac{1}{u_{0}}\, \delta^{\mu}_{0}$, and we get $\partial_{0} u_{\mu}=0$.

\vspace{0.5cm}

Using the above results, we can also show that
\begin{equation}
u^{\mu}_{~~;\mu}=0,~~\Gamma^{\mu}_{\mu
\nu}=\gamma^{\mu}_{\mu\nu},
\end{equation}
which we need in calculating the Riemann and Ricci tensors of (\ref{godel}). The Ricci tensor of $(g,\Gamma)$ is found as
\begin{equation}
R_{\mu \nu} = \bar{r}_{\mu\nu} + \frac{1}{2} \, f_{\mu}\,^{\alpha} \,
f_{\nu \alpha} + \frac{1}{2}\,(u_{\mu}\, j_{\nu} + u_{\nu}\,
j_{\mu}) +\frac{1}{4}\,f^2 \, u_{\mu} \, u_{\nu} \, , \label{ric}
\end{equation}
where $f^2 \equiv f^{\alpha \beta}\,f_{\alpha \beta}$,
\begin{equation}
j_{\mu} \equiv f^{\alpha}\,_{\mu \vert \alpha}=\nabla_{\alpha}\,
f^{\alpha}\,\,_{\mu}-{1 \over 2}\, f^2\, u_{\mu},
\end{equation}\label{jey}
and $\bar{r}_{\mu \nu}$ is the Ricci tensor of $\gamma^{\mu}_{\alpha\beta}$. The Ricci scalar is obtained as
\[ R = \bar{r} + \frac{1}{4}\, f^2 + u^{\mu} \, j_{\mu} \; , \]
where $\bar{r}$ denotes the Ricci scalar of $\bar{r}_{\alpha\beta}$.

\vspace{0.3cm}

\noindent {\bf Proposition 1}. {\it Let $(M,g)$ be a stationary
spacetime geometry described by the G\"{o}del-type metric (\ref{godel}) with $h_{\mu \nu}$ being the metric tensor of a $(D-1)$-dimensional (locally Euclidean) space. Then the Einstein tensor becomes
\begin{eqnarray}
G_{\mu\nu} & = & \bar{r}_{\mu\nu} - \frac{1}{2} \, h_{\mu\nu} \, \bar{r} +
\frac{1}{2}\, T^{f}_{\mu \nu} +
\frac{1}{2}\,(j_{\mu}\,u_{\nu}+j_{\nu}\,u_{\mu})\nonumber\\
&&+\left (
\frac{1}{4}\, f^{2} + \frac{1}{2} \, \bar{r} \right) u_{\mu} \,
u_{\nu}
- \frac{1}{2}\,(u^{\alpha}\,j_{\alpha})\,g_{\mu \nu} \, ,
\label{ein}
\end{eqnarray}
where $T^{f}_{\mu \nu}$ denotes the Maxwell energy-momentum tensor
for $f_{\mu\nu}$, i.e.
\begin{equation}\label{}
  T^{f}_{\mu \nu}\equiv f_{\mu\alpha}f_\nu^{~\alpha}-\frac{1}{4}\,g_{\mu\nu}f^2.
\end{equation}
}

\vspace{0.3cm}

\noindent This gives the Einstein tensor of the metric (\ref{godel}) in its full generality. To have a physical energy-momentum distribution, such as charged perfect fluid energy-momentum tensor, we shall assume that the $(D-1)$-space is an Einstein space and $j_{\mu}=0$, which yields the following proposition.

\vspace{0.3cm}

\noindent {\bf Proposition 2}. {\it Let $(M,g)$ be a stationary
spacetime geometry with the G\"{o}del type metric (\ref{godel}).
Let $h_{\mu \nu}$ be the metric tensor of a $(D-1)$-dimensional
Einstein space, i.e. $\bar{r}_{\mu \nu}={\bar{r} \over D-1} \, h_{\mu \nu}$, and let
$j_{\mu}=0$. Then the metric $g_{\mu \nu}$ satisfies the  Einstein
field equations with a charged perfect fluid
\begin{equation}
G_{\mu\nu}  \equiv  \frac{1}{2}\, T^{f}_{\mu \nu} + (p+\mathcal{E})\, u_{\mu}
\, u_{\nu} +p g_{\mu \nu}\, , \label{ein1}
\end{equation}
with
\begin{eqnarray}
&&\nabla_{\mu}\, f^{\mu \nu}-{1 \over 2}\, f^2 u^{\nu}=0, \label{eqn06}\\
&&p\equiv{(3-D)\over 2(D-1)}\, \bar{r},\\
&&\mathcal{E}\equiv{1 \over 4} f^2+{1 \over 2}\,\bar{r}.\label{rho}
\end{eqnarray}
Here $p$ is the pressure and $\mathcal{E}$ is the energy density of the fluid.}

\vspace{0.3cm}

\noindent {\bf Remark 1}. {\it When $D=3$  the 2-dimensional space
with the metric $h_{\mu \nu}$ is identically an Einstein space.
The only field equations to be solved are those given in
(\ref{eqn06}).}

\vspace{0.3cm}

The rest of the paper is organized as follows. In Sec. 2, we give a brief review of Einstein-Aether theory. Here we study the theory in generic $D$ dimensions and present the equations of motion. In Sec. 3, we show that the G\"{o}del-type metrics discussed in the introduction exactly solve the field equations of  the Einstein-Aether theory and the only equations that need to be solved are the $(D-1)$-dimensional Euclidean vacuum field equations and $(D-1)$-dimensional Euclidean Maxwell equations. In Sec. 4, we show how to construct such solutions in the theory by giving a recipe and present three explicit examples. Finally, we conclude our main results in Sec. 5.

\section{Einstein-Aether Theory}


The so-called Einstein-Aether theory \cite{jm,jac} is a vector-tensor theory of gravity in which the Lorentz symmetry is broken by the existence of a preferred frame of reference established by a vector field with fixed norm. The vector field defined in this way is referred to as the \textquotedblleft aether" and dynamically couples to the metric tensor of spacetime. The theory has attracted a lot of interest in recent years and investigated from various respects. For example, the stability issue of the aether was discussed in \cite{cdgt,dj}, time-independent spherically symmetric solutions and black hole solutions were analyzed in \cite{ej1,ej2,gej,tm,bjs,gs,bs,dww}, generalizations and cosmological implications were studied in \cite{cl,zfs1,bdfsz,zfs2,zzbfs}, and extensions to include other fields were considered in \cite{bl,ab}.

The theory is described by, in the absence of matter fields, the action
\begin{eqnarray} 
&&I={1 \over 16 \pi G}\, \int d^{D}\,x\sqrt{-g}\,{\cal L},\\
&&\nonumber\\
&&{\cal L}=R-2\Lambda-K^{\alpha \beta}\,\,_{\mu \nu}\, \nabla_{\alpha}\,
v^{\mu}\, \nabla_{\beta}\, v^{\nu}+\lambda\, (v^{\mu}\,
v_{\mu}+\varepsilon),\label{lag}\\
&&\nonumber\\
&&K^{\mu \nu}~_{\alpha \beta}=c_{1}\, g^{\mu \nu}\, g_{\alpha
\beta}+c_{2}\, \delta^{\mu}_{\alpha}\,
\delta^{\nu}_{\beta}+c_{3}\, \delta^{\mu}_{\beta}\,
\delta^{\nu}_{\alpha}-c_{4}\, v^{\mu}\, v^{\nu}\, g_{\alpha
\beta},\label{Ktensor}
\end{eqnarray}
where $\Lambda$ is the cosmological constant and $v^\mu$ is the aether field with the fixed-norm constraint
\begin{equation}\label{con}
v^{\mu}\, v_{\mu}=-\varepsilon,~~(\varepsilon=0,\pm1)
\end{equation}
which is enforced into the theory by the Lagrange multiplier $\lambda$ in (\ref{lag}). In general, for $\varepsilon=+1$ ($\varepsilon=-1$) the aether field will be timelike (spacelike), and for $\varepsilon=0$ it will be null. However, in this paper we shall consider only the case $\varepsilon=+1$ (a unit timelike vector field)--the case for which the Einstein-Aether theory is defined, actually. In a later communication \cite{GS}, we shall consider the case $\varepsilon=0$ (a null vector field) independently. The constants $c_{1}, c_{2}, c_{3}$ and $c_{4}$ appearing in (\ref{Ktensor}) are the dimensionless parameters of the theory and constrained by some theoretical and observational arguments \cite{jm,jac,cl,ej3,ems,fj,jac2,zfz,ybby,ybyb}.

Variation of this action with respect to $\lambda$ yields the constraint equation (\ref{con}), and variation with respect to the metric $g^{\mu\nu}$ and the aether $v^\mu$ gives respectively the field equations
\begin{eqnarray}
&&G_{\mu \nu}+\Lambda g_{\mu\nu}=\nabla_{\alpha}\, [ J^{\alpha}~_{(\mu}
\,v_{\nu)}-J_{(\mu}~^{\alpha}\, v_{\nu)}+J_{(\mu \nu )}\,
v^{\alpha}] \nonumber\\
&&~~~~~~~~~~~~~~~~~~+c_{1}(\nabla_{\mu}\, v_{\alpha}\, \nabla_{\nu}\,
v^{\alpha}-\nabla_{\alpha}\, v_{\mu}\, \nabla^{\alpha}\,
v_{\nu}) \nonumber \\
&&~~~~~~~~~~~~~~~~~~+c_{4}\, \dot{v}_{\mu}\, \dot{v}_{\nu}+\lambda v_{\mu}\,
v_{\nu}-{1 \over 2} L\, g_{\mu \nu}, \label{eqn01}\\
&&\nonumber\\
&&c_{4}\, \dot{v}^{\alpha}\, \nabla_{\mu}\,
v_{\alpha}+\nabla_{\alpha} \, J^{\alpha}~_{\mu}+\lambda
v_{\mu}=0,\label{eqn02}
\end{eqnarray}
where $\dot{v}^{\mu}\equiv v^{\alpha}\, \nabla_{\alpha}\, v^{\mu}$ and
\begin{eqnarray}
&&J^{\mu}~_{\nu}=K^{\mu \alpha}~_{\nu \beta}\, \nabla_{\alpha}\,
v^{\beta},\\
&&L=K^{\mu \nu}~_{\alpha \beta}\, \nabla_{\mu}\, v^{\alpha}\,
\nabla_{\nu}\, v^{\beta}.
\end{eqnarray}
In (\ref{eqn01}), we eliminated the term related to the constraint (\ref{con}). Multiplying the aether equation (\ref{eqn02}) by $v^\mu$, one can also derive
\begin{equation}\label{}
\lambda=c_{4}\, \dot{v}^{\alpha}\, \dot{v}_{\alpha}+v^{\alpha}\,
\nabla_{\beta}\, J^{\beta}~_{\alpha}.
\end{equation}

A special case of the theory (\ref{lag}) is the Einstein-Maxwell theory with dust distribution (no pressure) \cite{jm} (see also \cite{bek,zfs3}). Let $c_{2}=c_{4}=0$ and $c_{3}=-c_{1}$. Then the action becomes
\begin{equation}\label{ozel1}
I={1 \over 16 \pi G}\, \int d^{D}\,x\sqrt{-g}\,\left[R-2\Lambda-\frac{c_{1}}{2}\, F^2+\lambda\,
(v^{\mu}\, v_{\mu}+1)\right] \,
\end{equation}
with the field equations
\begin{eqnarray}
&&G_{\mu \nu}+\Lambda g_{\mu\nu}=c_{1}\, T^F_{\mu \nu}-\lambda v_{\mu}\,v_{\nu}, \label{ozel2}\\
&&\nabla_{\mu}\, F^{\mu \nu}= -{\lambda \over c_{1}} \, v^{\nu},\label{ozel3}\\
&&v^{\mu}\, v_{\mu}=-1, \label{son}
\end{eqnarray}
where $F_{\mu \nu}\equiv v_{\nu,\mu}-v_{\mu,\nu}$, $F^2\equiv F_{\mu\nu}F^{\mu\nu}$, and $T^{F}_{\mu \nu}$ is the Maxwell energy-momentum tensor for $F_{\mu \nu}$. Actually, this theory differs from the usual Einstein-Maxwell theory due to the last condition (\ref{son}) which brakes the gauge invariance of the theory. A generalization of this special theory is given in \cite{bek}, called TeVeS, which contains also a scalar (dilaton) field coupling to the unit timelike vector field and the metric tensor.

\section{G\"{o}del-Type Metrics in Einstein-Aether Theory}

Now we will show that G\"{o}del-type metrics with nonflat  background in arbitrary dimensions constitute solutions to Einstein-Aether theory. Let us first state the following proposition.

\vspace{0.3cm}

\noindent {\bf Proposition 3}. {\it If we assume that the timelike vector $v^{\mu}$  in the Einstein-Aether theory defined in the previous section be also a Killing vector, which satisfies the relations
\begin{equation}
v^{\mu}\,\,_{;\mu}=0,~~ \dot{v^{\mu}}=0,~~ v_{\nu;\mu}={1 \over 2}
F_{\mu \nu},
\end{equation}
where $F_{\mu \nu}\equiv v_{\nu,\mu}-v_{\mu,\nu}$, then we obtain
\begin{eqnarray}
J^{\mu}\,_{\nu}&=&{1 \over 2}\, (c_{1}-c_{3})\, F^{\mu}\,_{\nu}, \label{eqn03}\\
\lambda&=&-{1 \over 4}\, (c_{1}-c_{3})\, F^2, \label{eqn04}\\
L&=&{1 \over 4}\, (c_{1}-c_{3})\, F^2, \label{eqn05}
\end{eqnarray}
and the field equations (\ref{eqn01}) and (\ref{eqn02}) reduce
\begin{equation}\label{EinF}
G_{\mu \nu}+\Lambda g_{\mu \nu}=(c_{1}-c_{3})\left[{1 \over 2}\,T^{F}_{\mu \nu}+{1 \over
4} F^{2}\, v_{\mu}\,v_{\nu}\right],
\end{equation}
\begin{equation}\label{maxeqn2}
\nabla_{\alpha}\, F^{\alpha \mu}-{1 \over 2}\,
F^2\, v^{\mu}=0.
\end{equation}
Here $F^2\equiv F_{\mu\nu}F^{\mu\nu}$ and $T^{F}_{\mu \nu}$ is the energy-momentum tensor for the field
$F_{\mu \nu}$.}

\vspace{0.3cm}

\noindent An immediate consequence of this proposition is that the energy-momentum tensor of the aether field $v^\mu$ takes exactly the same form as that of a charged dust, as it can be seen readily from the right hand side of (\ref{EinF}). So this is indeed in the form desired in Proposition 2. Moreover, the aether equation (\ref{maxeqn2}) resembles the condition (\ref{eqn06}) on the vector field $u^\mu$ in G\"{o}del type of metrics. All these suggest that we can assume the scalings\footnote{These scalings were initially intended to get rid of the constraint $c_1-c_3=1$ and also have solutions with nonzero cosmological constant. However, at the end, it turned out that this is not possible (see Proposition 5 below).}
\begin{equation}\label{scale}
  v_\mu=\sigma u_\mu,~~g_{\mu\nu}=w^2g^{godel}_{\mu\nu},
\end{equation}
where $u^\mu$ is the unit timelike vector in the G\"{o}del-type metric (\ref{godel}), and $\sigma$ and $w$ are constant scale factors such that $\varepsilon\equiv\sigma^2/w^2=1$. This last identity stems from the fixed-norm constraint (\ref{con}). Therefore we have the proposition.

\vspace{0.3cm}

\noindent {\bf Proposition 4}. {\it If the relations (\ref{scale}) hold, we find that
\begin{eqnarray}
J^{\mu}\,_{\nu}&=&{1 \over 2\sigma}\, (c_{1}-c_{3})\, f^{\mu}\,_{\nu}, \label{eqn03}\\
\lambda&=&-{1 \over 4\sigma^2}\, (c_{1}-c_{3})\, f^2, \label{eqn04}\\
L&=&{1 \over 4\sigma^2}\, (c_{1}-c_{3})\, f^2, \label{eqn05}
\end{eqnarray}
where $f_{\alpha \beta} \equiv u_{\beta, \, \alpha} - u_{\alpha, \, \beta}$ and $f^2\equiv f_{\mu \nu}\, f^{\mu \nu}$, and the Einstein-Aether field
equations (\ref{EinF}) and (\ref{maxeqn2}) become
\begin{equation}\label{Einf}
G_{\mu \nu}+\Lambda g_{\mu \nu}=G^{godel}_{\mu\nu}+\Lambda \sigma^2 g^{godel}_{\mu \nu}=(c_{1}-c_{3})\left[{1 \over 2}\,T^{f}_{\mu \nu}+{1 \over
4} f^{2}\, u_{\mu}\,u_{\nu}\right],
\end{equation}
\begin{equation}\label{}
\nabla_{\alpha}\, F^{\alpha \mu}-{1 \over 2}\,
F^2\, v^{\mu}= \nabla_{\alpha}\, f^{\alpha \mu}-{1 \over 2}\, f^2\, u^{\mu}=0.
\end{equation}
Here $T^{f}_{\mu \nu}$ is the energy-momentum tensor for the field $f_{\mu \nu}$.}

\vspace{0.3cm}

\noindent Thus by comparing Proposition 4 with Proposition 2, we can draw the following conclusion:

\vspace{0.3cm}

\noindent {\bf Proposition 5}. {\it The stationary G\"{o}del-type
metrics in Proposition 2 solve the Einstein-Aether field equations (\ref{eqn01}) and (\ref{eqn02}) with
\begin{equation}
v_\mu=\sigma u_\mu,~~g_{\mu\nu}=\sigma^2g^{godel}_{\mu\nu},~~c_{1}-c_{3}=1,
\end{equation}
if and only if $h_{\mu \nu}$ is the metric of ($D-1$)-dimensional Ricci-flat
space (and hence $\Lambda=0$).}

\vspace{0.3cm}

At this point, it should be noted that in Proposition 2 it is assumed the ($D-1$)-dimensional metric $h_{\mu\nu}$ is an Einstein space and, as it turned out in Proposition 5, it is Ricci-flat. However, this assumption is actually unnecessary. Indeed, the field equations, together with the properties of the vector $u^\mu$, force $h_{\mu\nu}$ to be necessarily Ricci-flat. This can be seen as follows. Taking $j_\mu=0$ in (\ref{ein}) and using it in (\ref{Einf}) with $c_1-c_3=1$ produces the condition
\begin{equation}\label{conh}
\bar{r}_{\mu\nu}-\frac{1}{2}\bar{r}(h_{\mu\nu}-u_\mu u_\nu)+\Lambda \sigma^2g^{godel}_{\mu\nu}=0
\end{equation}
on the metric $h_{\mu\nu}$, which, with the help of the definition (\ref{godel}), becomes
\begin{equation}\label{conh1}
\bar{r}_{\mu\nu}-\frac{1}{2}(\bar{r}-2\Lambda \sigma^2)g^{godel}_{\mu\nu}=0.
\end{equation}
Now multiplying both sides by $u^\mu$ and using the fact that $u^\mu \bar{r}_{\mu\nu}=0$, one obtains
\begin{equation}\label{conh2}
  \bar{r}=2\Lambda \sigma^2.
\end{equation}
However, plugging this back into (\ref{conh1}) yields $\bar{r}_{\mu\nu}=0$ which in turn yields $\bar{r}=0$. Then from (\ref{conh2}) it necessarily turns out that $\Lambda=0$.

\vspace{0.3cm}

\noindent {\bf Remark 2}. {\it When $D=4$, the 3-dimensional background metric $h_{\mu \nu}$ becomes flat. In 3 dimensions Ricci flat spaces have zero curvature tensor. Hence, in four dimensions (D=4) we loose no generality by assuming $h_{\mu\nu}$ as a constant tensor ($\partial_{\alpha}\,h_{\mu\nu}=0$), as assumed in \cite{gur}.}

\vspace{0.3cm}

We can arrive at similar results for the special case of the Einstein-Aether theory described by the action (\ref{ozel1}); that is,

\vspace{0.3cm}

\noindent {\bf Proposition 6.} {\it The special Einstein-Aether field equations (\ref{ozel2}) and (\ref{ozel3}) are solved exactly by the stationary G\"{o}del type metrics in Proposition 2 with
\begin{equation}
v_{\mu}=\sigma u_{\mu},~~ g_{\mu \nu}=\sigma^2g^{godel}_{\mu \nu},~~c_1=\frac{1}{2},~~\lambda=-{1
\over 4\sigma^2}f^2,
\end{equation}
if and only if the (D-1)-dimensional space $h_{\mu\nu}$ is Ricci-flat (and hence $\Lambda=0$).
}

\vspace{0.3cm}

\section{Some Exact Solutions}

Now using the results of the previous section, we shall explicitly construct exact G\"{o}del-type solutions to Einstein-Aether theory. Here is the recipe for doing this:
\begin{enumerate}
  \item First, choose an appropriate Euclidean $(D-1)$-dimensional background metric $h_{\mu\nu}$ that satisfies the Ricci-flatness condition,
  \begin{equation}\label{CI}
   \bar{ r}_{\mu\nu}=0.
  \end{equation}
  \item Second, choose an appropriate timelike vector field $u^\mu$ that satisfies the condition (\ref{eqn06}), i.e. $j_\mu=0$, or equivalently \cite{gur1},
  \begin{equation}\label{CII}
    \partial_\mu(\bar{h}^{\mu\alpha}\bar{h}^{\nu\beta}\sqrt{h}f_{\alpha\beta})=0,
  \end{equation}
  where $\bar{h}^{\mu\nu}$ is the ($D-1$)-dimensional inverse of $h_{\mu\nu}$, i.e. $\bar{h}^{\mu\alpha}{h}_{\alpha\nu}=\bar{\delta}^\mu_\nu$.
  \item Third, construct the G\"{o}del-type metric defined by
  \begin{equation}\label{godel1}
    g^{godel}_{\mu\nu}=h_{\mu\nu}-u_\mu u_\nu.
  \end{equation}
  \item Finally, take
  \begin{equation}\label{}
  v_\mu=\sigma u_\mu,~~g_{\mu\nu}=\sigma^2g^{godel}_{\mu\nu},~~c_{1}-c_{3}=1,
  \end{equation}
  for some constant scale factor $\sigma$, which constitutes an exact solution to the Einstein-Aether field equations (\ref{eqn01}) and (\ref{eqn02}).
\end{enumerate}

\subsection{Solutions with ($D-1$)-dimensional flat backgrounds}

Let the ($D-1$)-dimensional (Euclidean) background metric $h_{\mu\nu}$ be flat. Since $h_{0\mu}=0$, this means that $h_{ij}=\bar{\delta}_{ij}$ with the Latin indices running from 1 to $(D-1)$. Thus our background reads
\begin{equation}\label{}
  ds^2_{D-1}=h_{\mu\nu}dx^\mu dx^\nu=\bar{\delta}_{ij}dx^i dx^j=(dx^1)^2+(dx^2)^2+\ldots+(dx^{D-1})^2.
\end{equation}
This choice of the background trivially satisfies the Ricci-flatness condition (\ref{CI}).

Now let us assume that $u_i=Q_{ij}x^j$ where $Q_{ij}$ is an antisymmetric constant tensor, i.e. $\partial_0Q_{ij}=0$. With this choice, $f_{ij}\equiv\partial_iu_j-\partial_ju_i=-2Q_{ij}$ and the condition (\ref{CII}) is trivially satisfied. Then
\begin{equation}\label{}
  u_\mu dx^\mu=u_0dx^0+u_idx^i=dx^0+\frac{1}{2}Q_{ij}(x^jdx^i-x^idx^j)
\end{equation}
and the following metric solves the Einstein-Aether theory with $v_\mu=\sigma u_\mu$ and $c_{1}-c_{3}=1$,
\begin{equation}\label{}
  ds^2=\sigma^2ds^2_{godel},
\end{equation}
where
\begin{eqnarray}\label{}
  ds^2_{godel}&=&g^{godel}_{\mu\nu}dx^\mu dx^\nu=(h_{\mu\nu}-u_\mu u_\nu)dx^\mu dx^\nu \nonumber\\
  &=&\bar{\delta}_{ij}dx^i dx^j-(u_\mu dx^\mu)^2 \nonumber\\
  &=&(dx^1)^2+(dx^2)^2+\ldots+(dx^{D-1})^2\nonumber\\
  &&~~~~~~~~~~~~-\left[dx^0+\frac{1}{2}Q_{ij}(x^jdx^i-x^idx^j)\right]^2.\label{hflat}
\end{eqnarray}
Specifically when $D=4$ (so $i,j=1,2,3$) and $Q_{13}=Q_{23}=0$ but $Q_{12}\neq 0$, with the coordinates labeled by $(t,x,y,z)$, the G\"{o}del-type metric (\ref{hflat}) becomes
\begin{equation}\label{}
  ds^2_{godel}=dx^2+dy^2+dz^2-\left[dt+\frac{1}{2}Q_{12}(ydx-xdy)\right]^2,
\end{equation}
or in cylindrical coordinates $(\rho,\phi,z)$,
\begin{equation}\label{}
  ds^2_{godel}=d\rho^2+\rho^2d\phi^2+dz^2-\left[dt-\frac{1}{2}Q_{12}\rho^2d\phi\right]^2.
\end{equation}
As is already discussed in \cite{gur,gur1,gur3}, since $\phi$ is a periodic variable, this spacetime contains closed timelike and null curves $x^\mu=(t,\rho,\phi,z)$ with $t,\rho,z=const.$ and $\rho\geq2/|Q_{12}|$.

The energy density (\ref{rho}) for the spacetime (\ref{hflat}) can be found to be
\begin{equation}\label{}
  \mathcal{E}\equiv{1\over 4}f^2={1\over 4}f_{ij}f^{ij}=Q_{ij}Q^{ij}=const.,
\end{equation}
which is regular everywhere in any dimension.

\subsection{Solutions with ($D-1$)-dimensional conformally flat backgrounds}

This time let us assume that the background metric $h_{ij}$  be conformally flat; that is, $h_{ij}=e^{2\psi(r)}\bar{\delta}_{ij}$, where $r$ is the radial distance in ${\mathbb{R}}^{D-1}$ defined by $r^2=\bar{\delta}_{ij}x^ix^j=x_ix^i$. Then, denoting the derivative with respect to $r$ by a prime, we work out  that (see, for instance, \cite{wald})
\begin{eqnarray}\label{rij}
&&\bar{r}_{ij}=-\bar{\delta}_{ij}\left[\psi''+(2D-5)\frac{\psi'}{r}+(D-3)\psi'^2\right]\nonumber\\
&&~~~~~~~~~~~~~~~~~~~~~~~~-x_ix_j\frac{D-3}{r}\left(\psi''-\frac{\psi'}{r}-\psi'^2\right).
\end{eqnarray}
For this to be equal to zero, we first assume that
\begin{equation}\label{}
  \psi''-\frac{\psi'}{r}-\psi'^2=0
\end{equation}
which has the generic solution
\begin{equation}\label{psi}
  \psi(r)=a-\ln(r^2+b)
\end{equation}
for some real constants $a$ and $b$. Putting (\ref{psi}) back into (\ref{rij}) produces
\begin{equation}\label{}
  \bar{r}_{ij}=\bar{\delta}_{ij}\frac{4b(D-2)}{(r^2+b)^2},
\end{equation}
which says that the Ricci-flatness condition (\ref{CI}) is achieved only when $b=0$ (since $D>2$ always).

Next we have to find an appropriate vector field $u^\mu$ that satisfies the condition (\ref{CII}). For this purpose, let us take $u_i=s(r)Q_{ij}x^j$ where again $Q_{ij}$ is an antisymmetric constant tensor. Then $f_{ij}$ turns out to be
\begin{equation}\label{fconflat}
  f_{ij}=2\,\frac{s'}{r}\,x^mQ_{m[i}x_{j]}-2sQ_{ij},
\end{equation}
where the square brackets denote antisymmetrization, and the condition (\ref{CII}) becomes
\begin{equation}\label{}
  \left(\frac{A}{r^2}\right)^{D-5}\left[s''+(10-D)\frac{s'}{r}+4(5-D)\frac{s}{r^2}\right]\bar{\delta}^{jl}Q_{lm}x^m=0,
\end{equation}
where $A\equiv e^a$. This equation is satisfied only if the expression inside the square brackets vanishes, which gives the generic solution
\begin{equation}\label{}
  s(r)=C_1r^{D-5}+\frac{C_2}{r^4}
\end{equation}
for some real constants $C_1$ and $C_2$. Therefore, we satisfy the condition (\ref{CII}), if we choose
\begin{equation}\label{}
  u_i=\left(C_1r^{D-5}+\frac{C_2}{r^4}\right)Q_{ij}x^j.
\end{equation}
Then the line element
\begin{eqnarray}\label{}
  ds^2_{godel}&=&g^{godel}_{\mu\nu}dx^\mu dx^\nu=(h_{\mu\nu}-u_\mu u_\nu)dx^\mu dx^\nu \nonumber\\
  &=&e^{2\psi}\bar{\delta}_{ij}dx^i dx^j-(u_\mu dx^\mu)^2 \nonumber\\
  &=&\frac{A^2}{r^4}\left[(dx^1)^2+(dx^2)^2+\ldots+(dx^{D-1})^2\right]\nonumber\\
  &&~~~~~-\left[dx^0+\frac{1}{2}\left(C_1r^{D-5}+\frac{C_2}{r^4}\right)Q_{ij}(x^jdx^i-x^idx^j)\right]^2\label{hconflat}
\end{eqnarray}
solves the Einstein-Aether field equation (\ref{eqn01}) and (\ref{eqn02}) if
\begin{equation}\label{}
  v_\mu=\sigma u_\mu,~~ds^2=\sigma^2ds^2_{godel},~~c_{1}-c_{3}=1.
\end{equation}
In four dimensions (i.e. $D=4$), as in the previous solution, assuming $Q_{13}=Q_{23}=0$ but $Q_{12}\neq 0$, and using cylindrical coordinates, one can write (\ref{hconflat}) in the form \cite{gur1}
\begin{equation}\label{}
  ds^2_{godel}=\frac{A^2}{r^4}(d\rho^2+\rho^2d\phi^2+dz^2)-\left[dt-\frac{1}{2}\left(\frac{C_1}{r}+\frac{C_2}{r^4}\right)Q_{12}\rho^2d\phi\right]^2,
\end{equation}
where $r^2=\rho^2+z^2$. As in the previous case, this spacetime also contains closed timelike and null curves when $t,\rho,z=const.$

One can further compute the energy density (\ref{rho}) for the spacetime (\ref{hconflat}) by using (\ref{fconflat}); so it is just a matter of computation to show that
\begin{eqnarray}\label{}
  \mathcal{E}&\equiv&{1\over 4}f^2={1\over 4}\bar{h}^{ik}\bar{h}^{jl}f_{ij}f_{kl}={e^{-4\psi}\over 4}\bar{\delta}^{ik}\bar{\delta}^{jl}f_{ij}f_{kl}\nonumber\\
  &=&\frac{1}{2A^4}\biggl\{\bar{\delta}^{ij}Q_{ik}Q_{jl}\frac{x^kx^l}{r^2}\biggl[C_1(D-5)r^{D-1}-4C_2\biggr]^2\nonumber\\
  &&~~~~~~~~~~~~~~~~~~~~~~~~~~+2\bar{\delta}^{ik}\bar{\delta}^{jl}Q_{ij}Q_{kl}\biggl[C_1r^{D-1}+C_2\biggr]^2\biggr\}.
\end{eqnarray}
It is evident from this expression that there is no singularity at $r=0$, for $D>1$ always.

\subsection{Solutions with ($D-1$)-dimensional Euclidean Tangherlini backgrounds}

Now consider the ($D-1$)-dimensional Euclidean Tangherlini \cite{tang} solution as our background metric $h_{ij}$. Labeling the coordinates by $(t,r,\theta_1,\theta_2,\ldots,\theta_{D-3})$, so that $x^\mu=(x^0,t,r,\theta_1,\theta_2,\ldots,\theta_{D-3})$, we can write the metric as follows
\begin{equation}\label{Tang}
  ds^2_{D-1}=V(r)dt^2+\frac{dr^2}{V(r)}+r^2d\Omega^2_{D-3},
\end{equation}
where
\begin{equation}\label{Vr}
  V(r)=1-2m\,r^{4-D}~~(D\geq4)
\end{equation}
with $m$ being the constant \textquotedblleft mass" parameter and $d\Omega^2_{D-3}$ is the metric on the $(D-3)$-dimensional unit sphere. This metric solves the Euclidean Einstein equations in vacuum, so the Ricci-flatness condition (\ref{CI}) is automatically satisfied.

To fulfill the condition (\ref{CII}), let us assume that $u_i=u(r)\delta^t_i$ for which $f_{ij}=u'(\delta^r_i\delta^t_j-\delta^r_j\delta^t_i)$, where the prime denotes differentiation with respect to $r$. Then, with the help of the metric (\ref{Tang}), it reduces to $(r^{D-3}u')'=0$, and this can easily be integrated to give
\begin{equation}\label{ur}
  u(r)=\left\{\begin{array}{ll}
               a\,r^{4-D}+b & \mbox{when $D\geq5$,} \\
               a\ln r+b & \mbox{when $D=4$,}
             \end{array}\right.
\end{equation}
where $a$ and $b$ are two arbitrary constants, the latter of which can be gauged away and taken to be zero.

Therefore, we can construct the following solution to the Einstein-Aether theory with $v_\mu=\sigma u_\mu$ and $c_{1}-c_{3}=1$:
\begin{equation}\label{}
  ds^2=\sigma^2ds^2_{godel},
\end{equation}
where
\begin{eqnarray}\label{}
  ds^2_{godel}&=&g^{godel}_{\mu\nu}dx^\mu dx^\nu=(h_{\mu\nu}-u_\mu u_\nu)dx^\mu dx^\nu \nonumber\\
  &=&h_{ij}dx^i dx^j-(u_\mu dx^\mu)^2 \nonumber\\
  &=&V(r)dt^2+\frac{dr^2}{V(r)}+r^2d\Omega^2_{D-3}-\left[dx^0+u(r)dt\right]^2,\label{hTang}
\end{eqnarray}
with $V(r)$ and $u(r)$ given by (\ref{Vr}) and (\ref{ur}), respectively. Here the coordinate $x^0$ plays the role of the time coordinate. When $D=5$, after relabeling the angular coordinates, (\ref{hTang}) becomes
\begin{equation}\label{}
  ds^2_{godel}=\left(1-\frac{2m}{r}\right)dt^2+\frac{dr^2}{1-\frac{2m}{r}}+r^2(d\theta^2+\sin^2\theta d\phi^2)-\left[dx^0+\frac{a}{r}dt\right]^2.
\end{equation}
As is discussed in \cite{gur1}, in this spacetime, only the region $r>2m$ is physical. Otherwise, the spacetime undergoes a signature change from $(-,+,+,+,+)$ to $(-,-,-,+,+)$.

We can also compute the energy density (\ref{rho}) for the spacetime (\ref{hTang}) to obtain
\begin{equation}\label{}
  \mathcal{E}\equiv{1\over 4}f^2=\left\{\begin{array}{ll}
               \frac{a^2(D-4)^2}{2}\,r^{2(3-D)} & \mbox{when $D\geq5$,} \\
               \frac{a^2}{2r^2} & \mbox{when $D=4$,}
             \end{array}\right.
\end{equation}
revealing the spacetime singularity at $r=0$ for $a\neq0$.

\section{Conclusion}

In this work, we showed that the G\"{o}del-type metrics \cite{gur1,gur2} in $D$ dimensions with nonflat background metric $h_{\mu \nu}$ solve the field equations of the Einstein-Aether theory. We proved that the  G\"{o}del-type metrics reduce the complete field equations of the Einstein-Aether theory to a $(D-1)$- dimensional vacuum field equations for the background metric and $(D-1)$- dimensional Maxwell equations corresponding to the unit timelike vector field, where all parameters of the theory are left free except  $c_{1}-c_{3}=1$. We also showed that G\"{o}del-type metrics in $D$ dimensions with nonflat background solve a special reduction of the Einstein-Aether theory \cite{jm,bek,zfs3}. In four dimensions with flat background, the results of this work coincide with those published in \cite{gur}. We presented some particular exact solutions to the Einstein-Aether theory in arbitrary dimensions: We considered $(D-1)$-dimensional Euclidean flat, conformally flat, and Tangherlini background metrics and showed that in the first two cases there are solutions allowing for the existence of closed timelike and closed null curves in the relevant spacetimes.

\section*{Acknowledgements}

We thank the referee for carefully reading our manuscript and providing constructive comments. This work is partially supported by the Scientific and
Technological Research Council of Turkey (TUBITAK).


\begin{thebibliography}{EMG}

\bibitem{gur1} M. G{\" u}rses, A. Karasu, and {\" O}. Sar{\i}o{\~ g}lu,
Class. Quantum Grav. {\bf 22}, 1527 (2005).
\bibitem{gur2} M. G{\" u}rses and {\" O}. Sar{\i}o{\~ g}lu,
Class. Quantum Grav. {\bf 22}, 4699 (2005).

\bibitem{godel} K. G\"{o}del, Rev. Mod. Phys. {\bf 21}, 447 (1949).

\bibitem{gur3} R. J. Gleiser, M. G{\" u}rses, A. Karasu, and {\" O}. Sar{\i}o{\~ g}lu,
Class. Quantum Grav. {\bf 23}, 2653 (2006).

\bibitem{gur4} M. G{\" u}rses, Gen. Rel. Grav. {\bf 42}, 1413 (2010).

\bibitem{gur} M. G{\" u}rses, Gen. Rel. Grav. {\bf 41}, 31 (2009).

\bibitem{jm} T. Jacobson and D. Mattingly, Phys. Rev. D {\bf
64}, 024028 (2001).
\bibitem{jac} T. Jacobson, Proc. Sci. QG-PH (2007) 020 [arXiv:0801.1547].


\bibitem{cdgt} S. M. Carroll, T. R. Dulaney, M. I. Gresham, and H. Tam, Phys. Rev. D {\bf 79}, 065011 (2009).
\bibitem{dj} W. Donelly and T. Jacobson, Phys. Rev. D {\bf 82}, 081501 (2010).

\bibitem{ej1} C. Eling and T. Jacobson, Class. Quantum Grav. {\bf 23}, 5625 (2006).
\bibitem{ej2} C. Eling and T. Jacobson, Class. Quantum Grav. {\bf 23}, 5643 (2006).
\bibitem{gej} D. Garfinkle, C. Eling, and T. Jacobson, Phys. Rev. D {\bf 76}, 024003 (2007).
\bibitem{tm} T. Tamaki and U. Miyamoto, Phys. Rev. D {\bf 77}, 024026 (2008).
\bibitem{bjs} E. Barausse, T. Jacobson, and T. P. Sotiriou, Phys. Rev. D {\bf 83}, 124043 (2011).
\bibitem{gs} C. Gao and Y. G. Shen, Phys. Rev. D {\bf 88}, 103508 (2013).
\bibitem{bs} E. Barausse and T. P. Sotiriou, Class Quantum Grav. {\bf 30}, 244010 (2013).
\bibitem{dww} C. Ding, A. Wang, and X. Wang, Phys. Rev. D {\bf 92}, 084055 (2015).

\bibitem{cl} S. M. Caroll and E. A. Lim, Phys. Rev. D {\bf 70}, 123525 (2004).
\bibitem{zfs1} T. G. Zlosnik, P. G. Ferreira, and G. D. Starkman, Phys. Rev. D {\bf 75}, 044017 (2007).
\bibitem{bdfsz} C. Bonvin, R. Durrer, P. G. Ferreira, G. D. Starkman, and T. G. Zlosnik, Phys. Rev. D {\bf 77}, 024037 (2008).
\bibitem{zfs2} T. G. Zlosnik, P. G. Ferreira, and G. D. Starkman, Phys. Rev. D {\bf 77}, 084010 (2008).
\bibitem{zzbfs} J. Zuntz, T. G. Zlosnik, F. Bourliot, P. G. Ferreira, and G. D. Starkman, Phys. Rev. D {\bf 81}, 104015 (2010).

\bibitem{bl} A. B. Balakin and J. P. S. Lemos, Ann. Phys. {\bf 350}, 454 (2014).
\bibitem{ab} T. Y. Alpin and A. B. Balakin, Int. J. Mod. Phys. D {\bf 25}, 1650048 (2016).

\bibitem{GS} M. G{\" u}rses and \c{C}. \c{S}ent\"{u}rk, arXiv:1604.02266.

\bibitem{ej3} C. Eling and T. Jacobson, Phys. Rev. D {\bf 69}, 064005 (2004).
\bibitem{ems} J. W. Elliot, G. D. Moore, and H. Stoica, JHEP {\bf 0508}, 066 (2005).
\bibitem{fj} B. Z. Foster and T. Jacobson, Phys. Rev. D {\bf 73}, 064015 (2006).
\bibitem{jac2} T. Jacobson, {\it Einstein-Aether Gravity: Theory
and Observational Constraints}, in the Proceedings of the {\bf
Meeting on CPT and Lorentz Symmetry (CPT 07)}, Bloomington,
Indiana, 8-11 Aug. 2007 [arXiv.0711.3822].
\bibitem{zfz} J. A. Zuntz, P. G. Ferreira, and T. G. Zlosnik, Phys. Rev. Lett. {\bf 101}, 261102 (2008).
\bibitem{ybby} K. Yagi, D. Blas, E. Barausse, and N. Yunes, Phys. Rev. D {\bf 89}, 084067 (2014).
\bibitem{ybyb} K. Yagi, D. Blas, N. Yunes, and E. Barausse, Phys. Rev. Lett. {\bf 112}, 161101 (2014).


\bibitem{bek} J. D. Bekenstein, Phys. Rev. D {\bf 70}, 083509 (2004).
\bibitem{zfs3} T. G. Zlosnik, P. G. Ferreira, and G.D. Starkman, Phys. Rev. D {\bf 74}, 044037 (2006).

\bibitem{wald} R. M. Wald, {\it General Relativity} (The University of Chicago Press, Chicago, 1984).

\bibitem{tang} F. R. Tangherlini, Nuovo Cim. {\bf 27}, 636 (1963).

\end{thebibliography}
\end{document}